\documentclass[12pt,aps,onecolumn,nobibnotes]{revtex4}

\usepackage{graphicx}%
\usepackage{dcolumn}
\usepackage{amsmath}
\usepackage{amssymb}

\makeatletter
\def\btt#1{\texttt{\@backslashchar#1}}%
\DeclareRobustCommand\bblash{\btt{\@backslashchar}}%
\makeatother

\newcommand{\be}{\begin{equation}}
\newcommand{\ee}{\end{equation}}
\newcommand{\bea}{\begin{eqnarray}}
\newcommand{\eea}{\end{eqnarray}}
\newcommand{\nn}{\nonumber}

\newcommand{\g}{GL(n, \mathbb{R})}
\newcommand{\gt}{GL(3, \mathbb{R})}
\newcommand{\son}{SO(n, \mathbb{R})}
\newcommand{\sot}{SO(3, \mathbb{R})}

\begin{document}

\title[Short Title]
{Classical mechanics on $GL(n, \mathbb{R})$ group and
Euler-Calogero-Sutherland model}
\author{A.M. Khvedelidze}
\thanks{Department of Theoretical Physics,
A. Razmadze Mathematical Institute, GE-380093 Tbilisi, Georgia.}
\author{\underline{D.M. Mladenov}}
\email{mladim@thsun1.jinr.ru}
\affiliation{
Bogoliubov Laboratory of Theoretical Physics,
Joint Institute for Nuclear Research, 141980 Dubna, Russia}

\begin{abstract}
Relations between the free motion on the $GL^+ (n, \mathbb{R})$ group manifold and
the dynamics of an $n$-particle system with spin degrees of freedom on a line interacting
with a pairwise $1/\sinh^2 x$ ``potential'' (Euler-Calogero-Sutherland model) is
discussed in the framework of Hamiltonian reduction.
Two kinds of reductions of the degrees of freedom are considered:
due to the continuous invariance and
due to the discrete symmetry.
It is shown that after projection on the corresponding invariant manifolds the resulting
Hamiltonian system represents the Euler-Calogero-Sutherland  model in both cases.
\end{abstract}

\maketitle


\section{Introduction}
\label{sec:level1}

In this contribution, we deal with two finite dimensional Hamiltonian
systems.
The first one is a generalization of the Calogero-Sutherland-Moser \cite{CSM} model
by introducing the internal degrees of freedom
\cite{GibbonsHermsen,Wojciechowski}
described by the Hamiltonian
\begin{equation} \label{eq:ES}
H_{ECS} = \frac{1}{2} \sum_{i=1}^N p_i^2 +
\frac{1}{8} \sum_{i\not=j}^N \frac{l_{ij}^2}{\sinh^2(x_i-x_j)}
\end{equation}
with canonical pairs $(x_i, p_i)$ obeying the nonvanishing Poisson brackets
\begin{equation}
\{ x_i, p_j \} = \delta_{ij} \,,
\end{equation}
and ``internal'' variables $l_{ab}$ satisfy the $\son$ Poisson bracket algebra
\begin{equation}
\{l_{ab}, l_{cd} \} =
 \delta_{bc} l_{ad} +
\delta_{ad} l_{bc} - \delta_{ac} l_{bd} -\delta_{bd} l_{ac}\,.
\end{equation}
The dynamics of the second system is given in terms of geodesic
motion on the $\g$ group manifold.
The corresponding Lagrangian based on the bi-invariant metric
on $\g$ is given by \cite{Arnold,Marsden}
\begin{equation}\label{eq:bilag}
L_{GL} = \frac{1}{2}\, \mbox{tr} \left( \dot g g^{-1} \right)^2 \,,
\end{equation}
where $ g \in \g $, and the dot over the symbols means differentiation with
respect to time.
Below we shall represent the Hamiltonian system
corresponding to this Lagrangian (\ref{eq:bilag}) in terms of a special
parameterization adapted to the action of the symmetry group of the system.
We demonstrate that the resulting Hamiltonian is a generalization of the
Euler-Calogero-Sutherland model (\ref{eq:ES}) with two types of internal
degrees of freedom.
Performing the Hamiltonian reduction owing to two types
of symmetry: continuous and discrete, we show how to arrive at the
conventional Hamiltonian of the Euler-Calogero-Sutherland model (\ref{eq:ES}).


\section{Bi-invariant geodesic motion on the group manifold}



\subsection{Explicit integration of the classical equation of motion}


The Euler-Lagrange equation following from the Lagrangian
(\ref{eq:bilag}) can be represented as
\begin{equation}
\frac{d}{dt}\left(g^{-1} \dot g \right)=0\,.
\end{equation}
This form demonstrates their explicit integrability
\begin{equation}
g(t) = g(0) \exp{(tJ)}
\end{equation}
with two arbitrary constant matrices $g(0)$ and $J$.


\subsection{Hamiltonian in terms of special coordinates}
\label{sec:level2}


The canonical Hamiltonian corresponding to the bi-invariant Lagrangian
(\ref{eq:bilag}) reads
\begin{equation}\label{eq:biham}
H_{GL} = \frac{1}{2}\, \mbox{tr} \left( \pi^T  g \right)^2 \,.
\end{equation}
The nonvanishing Poisson brackets between the
fundamental phase space variables are
\begin{equation}
\{g_{ab} \,, \pi_{cd}\} = \delta_{ab}\, \delta_{cd}\,.
\end{equation}

To find out the relation to the conventional
Euler-Calogero-Sutherland model (\ref{eq:ES}),
it is convenient to use the polar decomposition
\cite{Zelobenko} for an arbitrary element of $\g$.
For the sake of technical simplicity we investigate in details
the group $\gt$ hereinafter, i.e.
\begin{equation}\label{eq:polar}
g = O S\,,
\end{equation}
where $S$ is a positive definite $3 \times 3$ symmetric matrix, and
$O(\phi_1,\phi_2, \phi_3) = e^{\phi_1 J_3}e^{\phi_2 J_1}e^{\phi_3 J_3}$
is an orthogonal matrix with $SO(3, \mathbb{R})$ generators
$(J_a)_{ik} = \varepsilon_{iak}$.
Since the matrix $g$ represents an element of $\g$ group,
we can treat the polar decomposition (\ref{eq:polar}) as a
uniquely invertible transformation from the configuration variables
$g$ to a new set of six Lagrangian coordinates $S_{ij}$ and three coordinates $\phi_i$.
The induced transformation of momenta to new canonical pairs
$(S_{ab}, P_{ab})$ and $(\phi_a, P_a)$ is
\begin{equation}
\pi = O \left( P - k_a J_a \right)\,,
\end{equation}
where
\begin{equation}
k_a = \gamma^{-1}_{ab} \left(\eta^L_b -
\varepsilon_{bmn}\left(S P \right)_{mn}  \right)\,.
\end{equation}
Here $\eta^L_a$ are three left-invariant vector fields on $\sot$
\begin{eqnarray}
&& \eta^L_1 =
- \frac{ \sin\phi_3 }{ \sin\phi_2 }\,  P_1 -
\cos\phi_3 \,  P_2 +
\cot\phi_2 \sin\phi_3 \  P_3 \,,\nn\\ \label{eq:liv}
&& \eta^L_2 =
- \frac{ \cos\phi_3 }{ \sin\phi_2 }\,  P_1 +
\sin\phi_3 \,  P_2 +
\cot\phi_2 \cos\phi_3 \ P_3 \,,\\
&& \eta^L_3 =  - P_3\nn
\end{eqnarray}
and
\(
\gamma_{ik} = S_{ik} -  \delta_{ik} \ \mbox{tr}  S\,.
\)
In terms of the new variables, the canonical Hamiltonian takes the form
\begin{equation} \label{eq:hams}
H_{GL} =
\frac{1}{2} \mbox{tr} \left( PS \right)^2 +
\frac{1}{2} \mbox{tr} \left( J_a S J_b S \right) k_a k_b\,.
\end{equation}

\subsubsection{Restriction of the Hamiltonian to the Principal orbit}

The system (\ref{eq:hams}) is invariant under
the orthogonal transformations $S' = R^T\,S\,R$,
and the orbit space is given as a quotient space
${\mathcal S}/\sot$.
The quotient space ${\mathcal S}/\sot$ is a stratified manifold;
orbits with the same isotropy group are collected into {\it strata}
and uniquely parameterized by the
set of ordered eigenvalues of the matrix $S$ $x_1\leq x_2 \leq x_3$.
The strata are classified according to the isotropy groups
which are determined by the degeneracies of the matrix eigenvalues:
\begin{enumerate}
\item
{\it Principal orbit-type stratum},
when all eigenvalues are unequal $x_1< x_2 < x_3$, with
the smallest isotropy group $Z_2\otimes Z_2$\,.
\item
{\it Singular orbit-type strata}
forming the boundaries of orbit space with
\begin{enumerate}
\item two coinciding eigenvalues
(e.g. $x_1 = x_2$), when the isotropy group is $SO(2)\otimes Z_2$\,.
\item all three eigenvalues are equal
($x_1 = x_2 = x_3$),
here the isotropy group coincides with the isometry group $\sot$.
\end{enumerate}
\end{enumerate}

Now we shall at first restrict ourselves to the investigation of dynamics
which takes place on the {\it principal} orbits.
To write down the Hamiltonian describing the motion on the principal orbit stratum,
we introduce the coordinates along the slices $x_i$ and along the orbits $\chi$.
Namely, since the matrix $S$ is positive definite and symmetric,
we use the main-axes decomposition in the form
\begin{equation}\label{eq:mainaxes}
S = R^T(\chi) e^{2X} R(\chi)\,,
\end{equation}
where $R(\chi) \in \sot$ is an orthogonal matrix parameterized
by three Euler angles $\chi = (\chi_1, \chi_2, \chi_3)$,
and the matrix $e^{2X}$ is a diagonal
$e^{2X} = \mbox{diag}\, \| e^{2x_1}, e^{2x_2}, e^{2x_3} \|$.
The original physical momenta $P_{ik}$ are expressed in terms of the new canonical
pairs $(x_i, p_i)$ and $({\chi_i}, p_{\chi_i})$ as
\begin{equation} \label{eq:newmomenta}
P = R^T e^{- X}
\left(
\sum_{a=1}^{3}{\bar{\cal P}}_a {\bar\alpha}_a +
\sum_{a=1}^{3}{\cal P}_a {\alpha}_a
\right) e^{-X} R\,,
\end{equation}
with
\begin{eqnarray}
&& {\bar{\cal P}}_a = \frac{1}{2} p_a \,, \\
&&  {\cal P}_a  = - \frac{1}{4} \frac{\xi^R_a}{\sinh(x_b - x_c)}\,,
(\mbox{cyclic}\,\,\,\, \mbox{permutation} \,\,\, a\not=b\not= c)\,.
\end{eqnarray}
In the representation (\ref{eq:newmomenta}), we introduce the orthogonal basis
for the symmetric $3 \times 3$ matrices
$\alpha_A = ( \overline{\alpha}_i ,\ \alpha_i ) \  i = 1, 2, 3 $
with the scalar product
\begin{eqnarray}
\mbox{tr} (\bar\alpha_a\, \bar\alpha_b)= \delta_{ab}\,,
\quad
\mbox{tr} (\alpha_a\, \alpha_b)=2\delta_{ab}\,,
\quad
\mbox{tr} (\bar\alpha_a\, \alpha_b)= 0\nn
\end{eqnarray}
and the $\sot$ right-invariant Killing vectors
\begin{eqnarray}
&& \xi^R_1 = - p_{\chi_1}\,,\\
&& \xi^R_2 =
\sin\chi_1 \cot\chi_2 \ p_{\chi_1} -
\cos\chi_1 \  p_{\chi_2} -
\frac{\sin\chi_1}{\sin\chi_2}\ p_{\chi_3} \,,\\
&& \xi^R_3 =
- \,\,\cos\chi_1 \cot\chi_2 \ p_{\chi_1} -
\sin\chi_1 \  p_{\chi_2} +
\frac{\cos\chi_1}{\sin\chi_2}\ p_{\chi_3} \,.
\end{eqnarray}
After passing to these main-axes variables, the canonical Hamiltonian reads
\begin{equation}\label{eq:hamsma}
H_{GL} =
\frac{1}{8} \sum_{a=1}^{3} p_a^2 +
\frac{1}{16} \sum_{(abc)}  \frac{(\xi^R_a)^2}{\sinh^2(x_b - x_c)} -
\frac{1}{4} \sum_{(abc)}
\frac{\left(R_{ab} \eta^L_b + \frac{1}{2} \xi^R_a \right)^2}{\cosh^2(x_b - x_c)}
\,.
\end{equation}
Here $(abc)$ means cyclic permutations $a\neq b \neq c$.
Thus, the integrable dynamical system describing the free motion
on principal orbits represents, in the adapted basis, the
Generalized Euler-Calogero-Sutherland model.
The generalization consists in the introduction of two types of internal
dynamical variables
$\xi$ and $\eta$ --- ``spin'' and ``isospin'' degrees of freedom,
interacting with each other.
Below, the relations to the standard Euler-Calogero-Sutherland model
(\ref{eq:ES}) are demonstrated.

\subsubsection{Restriction of the Hamiltonian to the Singular orbit}
\label{sec:level3}

The motion on the Singular orbit is modified due to the presence
of a continuous isotropy group.
In the case of $\gt$, it is $SO(2)\otimes Z_2$.
Applying the same machinery as for the Principal orbits
to the two-dimensional orbit
$ (x_1 = x_2 = x\,\,, x_3 = y)$ \,,
one can derive the Hamiltonian
\begin{equation}\label{eq:2cs}
H^{(2)}_{GL} =
\frac{1}{8}
(p_x^2 + p_y^2) +\frac{g^2}{\sinh^2(x - y)} -
\frac{\overline{g}^2}{\cosh^2(x - y)}
\end{equation}
with two arbitrary constants $g^2$ and $\overline{g}^2$
related to the value of the spin $\xi$ and isospin $\eta$.
Due to the translation invariance, the equations of motion
are equivalent to the corresponding equations for a one-dimensional problem
and thus the system (\ref{eq:2cs}) is integrable.


\section{Reduction to Euler-Calogero-Sutherland model}



\subsection{Reduction using discrete symmetries}


Now we shall demonstrate how the II$A_3$ Euler-Calogero-Sutherland model
arises from the canonical Hamiltonian (\ref{eq:biham}) after projection onto a
certain invariant submanifold determined by discrete symmetries.
Let us impose the condition of symmetry of the matrices $g \in \g$
\begin{equation}\label{eq:primconst}
\chi_a^{(1)} = \varepsilon_{abc} g_{bc}=0 \,.
\end{equation}
In order to find an invariant submanifold, it is necessary to supplement the constraints
(\ref{eq:primconst}) with the new ones
\begin{equation}\label{eq:secconst}
\chi_a^{(2)} = \varepsilon_{abc} \pi_{bc}=0 \,.
\end{equation}
One can check that the surface defined by both constraints
(\ref{eq:primconst}) and (\ref{eq:secconst})
represents an invariant submanifold in the $\gt$ phase space, and
the dynamics of the corresponding induced system is governed by the
reduced Hamiltonian
\begin{equation}
H_{GL}\vert_{ \chi_a^{(1)} = 0,\,\,\, \chi_a^{(2)} = 0 } =
\frac{1}{2} \mbox{tr} \left( PS \right)^2\,.
\end{equation}
The matrices $S$ and $P$ are now symmetric nondegenerate matrices,
and one can be convinced that this expression
leads to the Hamiltonian of the II$A_3$ Euler-Calogero- Sutherland model.
To verify this statement, it is necessary to note that
after projection on the invariant submanifold,
the canonical Poisson structure is changed.
We have to deal with the new Dirac brackets
\begin{equation}
\{F, G\}_D = \{F, G\}_{PB} - \{F,\chi_a\} C_{ab}^{-1} \{\chi_b, G\}
\end{equation}
for arbitrary functions on the phase space.
In our case, because
\(
C_{ab} = \| \{ \chi_a^{(1)}, \chi_b^{(2)}\}\| = 2 \delta_{ab}\,,
\)
the fundamamental Dirac brackets between the main-axes variables are
\[ \{ x_a, p_b \}_D = \frac{1}{2} \delta_{ab} \,, \quad
\{ \chi_a, p_{\chi_b} \}_D = \frac{1}{2} \delta_{ab}
\]
and the Dirac bracket algebra for the right-invariant vector fields
on $\sot$ reduces to
\[
\{ \xi_a^R , \xi_b^R  \}_D = \frac{1}{2} \varepsilon_{abc} \xi_c^R\,.
\]
Thus, after rescaling of the canonical variables, one can be convinced that
the reduction via discrete symmetry indeed leads to the
II$A_3$ Euler-Calogero- Sutherland model.


\subsection{Reduction due to the continuous symmetry}


The integrals of motion corresponding to the geodesic motion with respect to the
bi-invariant metric on $\g$  group are
\begin{equation}
J_{ab} = (\pi^T g)_{ab}.
\end{equation}
The algebra of this integrals realizes on the symplectic level the
$\g$ algebra
\begin{equation}
\{J_{ab}, J_{cd}\} = \delta_{bc} J_{ad} - \delta_{ad} J_{cb}\,.
\end{equation}
After transformation to the scalar and rotational variables,
the expressions for $J$ reads
\begin{equation}\label{eq:intr}
J = \sum_{a=1}^3  R^T
 \left(p_a {\bar\alpha}_a - i_a \alpha_a - j_a J_a
\right)R\,,
\end{equation}
where
\begin{equation}\label{eq:int+-}
i_a = \frac{1}{2}\xi^R_a \coth(x_b - x_c) +
\left(R_{ab}\eta^L_b + \frac{1}{2}\xi^R_a \right) \tanh(x_b - x_c)
\end{equation}
and
\begin{equation}
j_a = R_{ab}\eta^L_b + \xi^R_a \,.
\end{equation}
When these integrals are used, there appear several ways to choose an invariant manifold
and to derive the corresponding reduced system.
Let us consider the surface on phase space defined by the constraints
\begin{equation} \label{con1}
\eta^R_a = 0 \,.
\end{equation}
These constraints, in the Dirac's terminology \cite{DiracL,HenTeit}
are first class constraints
\(
\{ \eta^R_a, \eta^R_b\}= -\epsilon_{abc}\eta^R_c
\)\,,
and the surface (\ref{con1}) is invariant under the evolution governed by the Hamiltonian
\[
\{ \eta^R_a, H_{GL}\}= 0\,.
\]
Using the relation between left and right-invariant Killing fields
\(
\eta^R_a= O_{ab}\eta^L_b
\)\,,
we find out that after projection to the constraint surface
(\ref{con1}), the Hamiltonian reduces to
\begin{equation}\label{eq:esham+}
H_{GL}(\eta^R_a= 0) =
\frac{1}{8} \sum_a^3 p_a^2 +
\frac{1}{4} \sum_{(abc)}  \frac{(\xi^R_a)^2}{\sinh^22(x_b - x_c)}
\,.
\end{equation}
After rescaling of the variables $2x_a \to x_a$, one is convinced that
the derived Hamiltonian coincides with the
Euler-Calogero-Sutherland Hamiltonian (\ref{eq:ES}),
where the intrinsic spin variables are $l_{ij} = \epsilon_{ijk}\xi^R_k $.
Note that performing the reduction to the surface defined by the
vanishing integrals $j_a = 0$,
we again arrive at the same Euler-Calogero-Sutherland system.


\subsection{Lax-pair for Generalized Euler-Calogero-Sutherland model}


The expressions (\ref{eq:intr}) for the integrals of motion
allow us to rewrite the classical equation of motion
for the Generalized Euler-Calogero-Sutherland model
in the Lax form

\begin{equation}
\dot{L}= [A, L ]\,,
\end{equation}
where $3 \times 3$ matrices are given explicitly as
\[
\begin{array}{lr}
L =
\left(
\begin{array}{ccc}
p_1       &  L^+_3,  &     L^-_2  \\
L^-_3,    &   p_2    &     L^+_1  \\
L^+_2,    &  L^-_1,  &     p_3    \\
\end{array}
\right)
\end{array}
\]

and

\[\label{eq:Lmatrix}
\begin{array}{lr}
A =
\frac{1}{4}\,e^{+X}
\left(
\begin{array}{ccc}
  p_1     & - A_3,     &   A_2  \\
  A_3,    &   p_2,     & - A_1  \\
- A_2,    &   A_1,     &   p_3    \\
\end{array}
\right)
e^{-X} \,,
\end{array}
\]
where
\begin{eqnarray}
&& L^\pm_1 =
- \frac{1}{2}\frac{\xi^R_1}{\sinh (x_2 - x_3)} \pm
\frac{ R_{1m} \eta^L_m + \frac{1}{2}\xi^R_1 }{ \cosh (x_2 - x_3) }
\,,\\
&& L^\pm_2 =
- \frac{1}{2}\frac{\xi^R_2}{\sinh(x_3 - x_1)} \pm
\frac{ R_{2m} \eta^L_m + \frac{1}{2}\xi^R_2 }{ \cosh (x_3 - x_1) }
\,,\\
&& L^\pm_3 =
- \frac{1}{2}\frac{\xi^R_3}{\sinh(x_1 - x_2)} \pm
\frac{ R_{3m} \eta^L_m + \frac{1}{2}\xi^R_3 }{ \cosh(x_1 - x_2) }
\end{eqnarray}
and
\begin{eqnarray}
&& A_1 =
\frac{1}{2}\frac{\xi^R_1}{\sinh^2 (x_2 - x_3)} -
\frac{ R_{1m} \eta^L_m + \frac{1}{2}\xi^R_1 }{ \cosh^2 (x_2 - x_3) }
\,,\\
&& A_2 =
\frac{1}{2}\,\frac{\xi^R_2}{\sinh^2 (x_3 - x_1)} -
\frac{ R_{2m} \eta^L_m + \frac{1}{2}\xi^R_2 }{ \cosh^2 (x_3 - x_1) }
\,,\\
&& A_3 =
\frac{1}{2}\,\frac{\xi^R_3}{\sinh^2 (x_1 - x_2)} -
\frac{ R_{3m} \eta^L_m + \frac{1}{2}\xi^R_3 }{ \cosh^2 (x_1 - x_2) }
\,.
\end{eqnarray}


\section {Concluding Remarks}


In this talk, we have discussed the generalization of the
Euler-Calogero-Sutherland model by introducing two internal variables
``spin'' and ``isospin'', using the integrable model based on the general
matrix group $\g$.
We outline its relation to the well-known integrable model.
Our consideration confirms once more that the clue to an
integrability of a model is often hidden in the possibility to connect it
with a known higher-dimensional exactly solvable system by its symplectic
reduction to its invariant submanifold \cite{Arnold,Marsden}.
A rich spectrum of these types of finite-dimensional models, obtained by the
generalized ``momentum map'' is well-known (see e.g. \cite{PerelBook}).
Over the last decade it has been recognized that the same happens in the
infinite-dimensional case.
Integrable two-dimensional field theories have
been found from the so-called WZNW theory applying the Hamiltonian reduction
method \cite{FO'RRTW}.
An important class of finite-dimensional systems was
discovered by the Hamiltonian reduction method from the so-called matrix
models (for a recent review see e.g.\cite{Polychronakos}).
The interest to this type of models has a long history starting with the
Wigner study of the
statistical theory of energy levels of complex nuclear system \cite{Mehta}.
Nowadays we have revival of the interest to a matrix models connected with
the search of relations between the supersymmetric Yang-Mills theory and
integrable systems (for a modern review see e.g. \cite{DPReview}).
The relation between the Euler-Calogero-Moser model and the $SU(2)$ Yang-Mills
theory in long-wavelength approximation was obtained in (\cite{ECM-YM}).

\begin{acknowledgments}
It is a pleasure to thank the local organizers of the
XXIII International Colloquium on
Group Theoretical Methods in Physics
and
B. Dimitrov, V.I. Inozemtsev, A. Kvinikhidze, M.D. Mateev, and
P. Sorba for illuminating discussions.

\end{acknowledgments}


\end{document}